# PERFORMANCE EVALUATION OF A NEW ROUTE OPTIMIZATION TECHNIQUE FOR MOBILE IP

Moheb r. Girgis[1], Tarek m. Mahmoud[1],
Youssef s. Takroni[1] and hassan s. Hassan[1]

[1]Computer Science Department, Minia University, El- Minia,61519, Egypt.

moheb@mail.minia.edu.eg , Tarek_2ms@yahoo.com , ysaad@mail.minia.edu.eg and hassan@mail.minia.edu.eg

## ABSTRACT

*Mobile ip (mip) is an internet protocol that allows mobile nodes to have continuous network connectivity to the internet without changing their ip addresses while moving to other networks. The packets sent from correspondent node (cn) to a mobile node (mn) go first through the mobile node's home agent (ha), then the ha tunnels them to the mn's foreign network. One of the main problems in the original mip is the triangle routing problem. Triangle routing problem appears when the indirect path between cn and mn through the ha is longer than the direct path. This paper proposes a new technique to improve the performance of the original mip during the handoff. The proposed technique reduces the delay, the packet loss and the registration time for all the packets transferred between the cn and the mn. In this technique, tunneling occurs at two levels above the ha in a hierarchical network. To show the effectiveness of the proposed technique, it is compared with the original mip and another technique for solving the same problem in which tunneling occurs at one level above the ha. Simulation results presented in this paper are based on the ns2 mobility software on linux platform. The simulations results show that our proposed technique achieves better performance than the others, considering the packet delay, the packet losses during handoffs and the registration time, in different scenarios for the location of the mn with respect to the ha and fas.*

## 1. INTRODUCTION

Today, the number of wireless and mobile devices connected to the Internet is strongly growing. Wireless links and networks, mobile users and mobility-related services form an increasing part of the Internet infrastructure. These wireless and mobile parts are normally connected to larger, wired networks. Mastering the key concepts in mobile, wireless and wired technology areas are therefore of increasing importance in the society of today. Mobile IP is an Internet protocol, defined by the Internet Engineering Task Force (IETF) that allows users keep the same IP address, and stay connected to the Internet while roaming between networks. The key feature of Mobile IP design is that all required functionalities for processing and managing mobility information are embedded in well-defined entities, the Home Agent (HA), Foreign Agent (FA), and Mobile Nodes (MNs) [1, 2]. When a MN moves from its Home Network (HN) to a Foreign Network (FN), the correct delivery of packets to its current point of attachment depends on the MN's IP address, which changes at every new point of attachment. Therefore, to ensure packets delivery to the MN, Mobile IP (MIP) allows the MN to use two IP addresses: The Home address and Care-of-Address (CoA). The home address is static and assigned to the MN at the HN; CoA on the other hand is dynamic and represents the current location of the MN [2].

Original MIP has many problems such as home agent faults tolerance [14], HA overloading, and triangle routing problem. Triangle routing problem is considered as one of the main problems facing the implementation of MIP. When a Corresponding Node (CN) sends traffic to the MN, the traffic gets first to the HA, which encapsulates this traffic and tunnels it to the FA. The FA de-tunnels the traffic and delivers it to the MN. The route taken by this traffic is triangular in nature, and the most extreme case of routing can be observed when the CN and the MN are in the same subnet [6, 7]. Many protocols have been invented to solve the triangle routing problem, such as forward tunneling and binding cache [2], bidirectional route optimization [8], the smooth handoff technique [11], a virtual home agent [12], and a port address translation based route optimization scheme [9]. Also, Kumar et al. [19] presented a route optimization technique in which the tunneling is done at one level above the HA in a hierarchical network instead of tunneling at the HA.

This paper proposes a new technique for solving the triangle routing problem in Mobile IP and reducing the delay which is not acceptable for real time applications such as Voice over IP (VoIP). The proposed technique reduces the packet delay, the packet loss and the registration time for all the packets transferred between the CN and the MN. In this technique, tunneling occurs at two levels above the HA in a hierarchical network instead of tunneling at the HA. To evaluate the performance of the proposed technique, a simulation is carried out using NS-2 simulator on Linux platform [16, 17, 18]. In this simulation the proposed technique is compared with the original MIP and the "one level up" technique proposed by Kumar et al. [19].

The rest of this paper is organized as follows. In section 2, we introduce the original MIP operations. The proposed technique used to improve the performance of the original MIP is described in section 3. Performance evaluation and experiments results are presented in section 4, followed by the conclusion in section 5.

2. Original MIP Operations

MIP supports mobility by transparently binding the home address of the MN with its CoA. Some specialized routers known as mobility agents maintain this mobility binding. Mobility agents are of two types - home agents and foreign agents. The HA is a designated router in the home network of the MN, which maintains the mobility binding in a mobility-binding table, as shown in Table 1. The purpose of this table is to map the mobile node's home address to its CoA and forward packets accordingly.

Table 1. Mobility-Binding Table

| MN's Home Address | MN's care-of-Address | Lifetime (sec) |
|---|---|---|
| 193.227.47.250 | 128.172.23.78 | 90 |
| 193.227.47.30 | 118.133.22.94 | 135 |
| ... | ... | ... |

Foreign agents are specialized routers on the foreign network where the MN is currently visiting. The foreign agent maintains a visitor list, which contains information about the mobile nodes currently visiting that network. Table 2 shows an instance of a visitor list.

Table 2. Visitor list

| MN's Home Address | MN's Home Agent Address | MAC Address | Lifetime (sec) |
|---|---|---|---|
| 193.227.47.250 | 193.227.47.100 | 00-00-21-00-95-c9 | 120 |
| 131.193.17.3 | 131.193.17.100 | 00-40-51-00-33-a9 | 160 |
| ... | ... | ... | ... |

The original MIP involves three operations: Agent discovery, Registration and Tunneling.

## 2.1 Agent Discovery

The MN is responsible for discovering whether the MN is in a home or foreign network. This process is done by either the Agent Advertisement or Agent Solicitation communication process [13]. Usually, the FA periodically broadcasts the Internet Router Discovery Protocol (IRDP) message [13] in its own network to let the visited MN know the FA is here and what services the FA provides (Agent Advertisement). Thus, the MN knows which network it belongs to. In case the MN does not receive this message, it can request the service by sending a solicitation message to inform the FA directly (Agent Solicitation). If there is no answer back during a limited time, the MN attempts to use the Dynamic Host Configuration Protocol (DHCP) to acquire a new IP address. [1, 13].

## 2.2 Mobility Agent with Registration

The purpose of the registration process in MIP is to inform the mobility agent with a mobile node's new IP address and update the binding information between home address of the mobile node and its CoA. This allows TCP connections to be maintained without a disruption and correspondent nodes to communicate with mobile nodes directly. Once a mobile node has a CoA, it registers its CoA with its home agent so that the home agent knows where to forward its packets as shown in Figure 1. Depending on the network configuration, the mobile node could either register directly or indirectly [1, 2]. The indirect registration steps are as follows:

   a) The mobile node sends a Registration Request to the prospective foreign agent to begin the registration process.
   b) The foreign agent processes the Registration Request and then relays it to the home agent.
   c) The home agent sends a Registration Reply to the foreign agent to permit or deny the request.
   d) The foreign agent processes the Registration Reply and sends it to the mobile node.

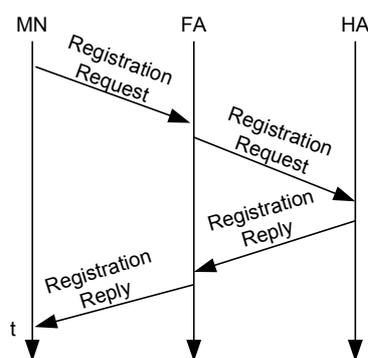

Figure 1. Indirect Registration

## 2.3 Mobility Agent with Tunneling

In order to deliver packets from the MN to the HA and vice versa, either the FA or the MN has to do tunneling to avoid the route propagation problem. After the tunnel is established, it is considered as just only one hop end-to-end from either the FA or the MN to the HA. Basically, there are three kinds of encapsulation technique. First, a traditional IP-in-IP encapsulation [3] simply encapsulates the original IP packet within the new IP header. Second, Perkins proposed the idea of the Minimal Encapsulation, used to avoid the repetition of the IP fields [4]. Third, The Generic routing encapsulation (GRE) [5] is another tunneling protocol, which supports various kinds of transport protocols over IP network.

## 3. Description of the Proposed Technique

This section describes our proposed technique to solve the triangle routing problem of MIP to reduce the handoff delay, packet loss and the registration time. A MN initiates a handoff whenever it enters into an area of a mobility agent different from its current area. During a handoff, MN is unreachable and packets may be lost if no buffering scheme is used.

A hierarchical network is considered for the new technique, in which the routers and nodes are arranged logically in the form of parents and children [15]. The network is divided into domains clubbed together to form higher level domains and so on till one reaches at the top, which is known as the root. The hierarchical network helps in organizing and managing the network. It represents an idealized Internet where optimal paths are used. Our hierarchical network consists of 4 levels contains the HA, FAs, MN, and standard IPv4 nodes without mobility support as illustrated in Figure 2. The addressing system used in this network model consists of 3 levels of hierarchy: domain.cluster.node, according to NS2 simulator. In the proposed technique as the MN moves away from its HN, it registers a new CoA with the HA. The HA forwards the new CoA and all information related to the MN to a router that has the same functionality as a mobile agent and resides two levels above the original HA, which we refer to as *surrogate HA (SHA)*. The packets destined for MN is tunneled at that router instead of the HA node, which saves the transmission time. Theses steps can be summarized as follows:

1. Mobility agents (HA and FA) advertise their presence using its agent advertise messages.
2. The MN receives an agent advertisement message and determines whether it is in the HN or in a FN
3. When MN detects that it is in a FN, it requests a CoA from this FN.
4. Then MN registers it's CoA with it's HA using the registration and replay messages.
5. The HA forwards the MN's CoA to the SHA
6. When packets are sent to the MN, they are intercepted by the SHA rather than the original HA, and then the SHA tunnels them to the MN current location.

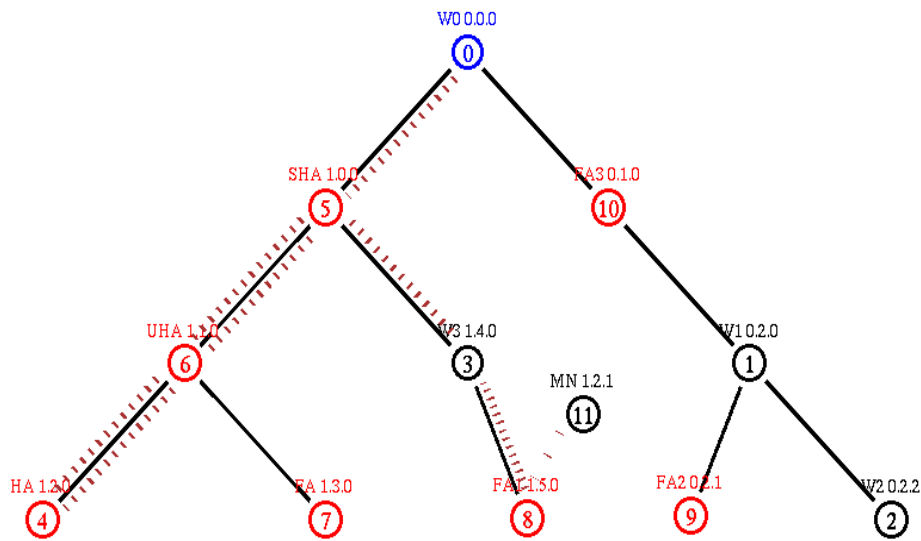

Figure 2. Traffic data flow path for the original MIP

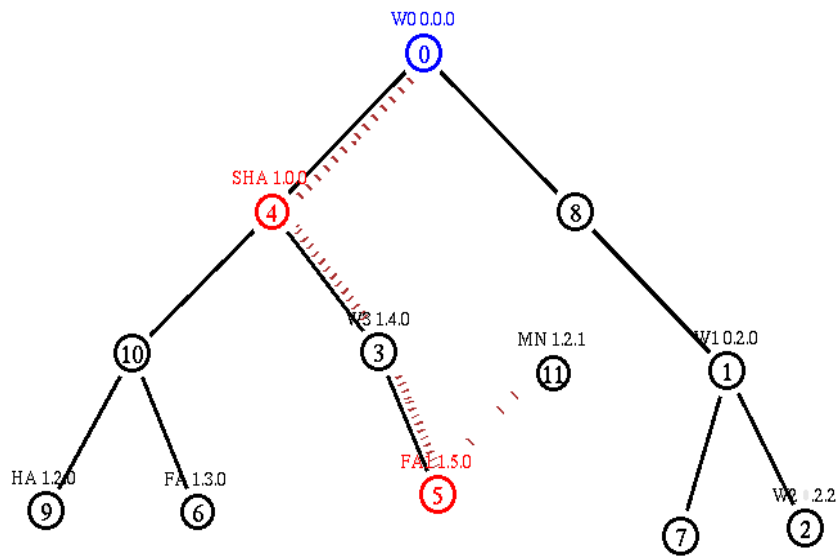

Figure 3. Traffic data flow path for the proposed technique

To illustrate the triangle problem and the proposed technique we consider the following example. Suppose that the MN will receive traffic from a correspondent node (W0) which is the root of the hierarchal tree with an address 0.0.0. If the MN (1.2.1) moves from its HA (1.2.0) to the FA (1.5.0), the original MIP path will be (0.0.0, 1.0.0, 1.1.0, 1.2.0, 1.1.0, 1.0.0, 1.4.0, 1.5.0) to the MN as illustrated in Figure 2. In the proposed technique, the new current location of the MN is forwarded to SHA (1.0.0), so the traffic data flow will take the path (0.0.0, 1.0.0, 1.4.0, 1.5.0), as illustrated in Figure 3. This overcomes the triangle routing problem of the original MIP and enhances the performance by reducing the delay, the packets loss and the registration time.

## 4. PERFORMANCE EVALUATION

In this section we present our simulation results of the original MIP, the "one level up", and the proposed technique. Simulations are conducted to investigate three critical performance issues, i.e. packet delay, packet loss during handoffs and registration time. The simulations are run using the open source NS2 simulator from Lawrence Berkeley National Laboratory (LBNL), which is widely used in the networking community to study IP networks. In the network model, it is assumed that all the routers have mobility support and the router one level and two levels above the home agent can keep the current position of mobile in its routing table. It is also assumed that the router one level and two levels above the home agent is capable of performing the encapsulation (IP in IP). The random mobility pattern's movement scenario for our simulations were generated using the setdest utility of NS2, with one MN, the grid area, the maximum speed, traffic type, link delay, transmitter range, bandwidth, packet rate, packet size, and simulation time as its parameters. The parameters values that have been used for the mobility simulations are summarized in Table 3.

Table 3. Parameters used during Mobility Simulations

| Parameter | Value |
|---|---|
| grid area | 1000 x 1000 m |
| maximum speed | 75 m/s |
| traffic type | Constant Bit Rate (CBR) |
| Link delay | 20 ms |
| transmitter range | 150 m |
| bandwidth | 2 Mb |
| packet rate | 5 Packet/sec |
| packet size | 200 byte |
| simulation time | 20 s |

Five simulation scenarios are considered to evaluate the performance of the considered techniques. Each scenario is based on the point of attachment and mobility of MN. The handoff delay, the packet loss and the registration time are calculated in each scenario for each technique, and the results are presented in the form of graphs and tables. For all simulation scenarios, the mobile traffic starts from 0.5 sec and the handoff occurs at about 16 sec and terminates at about 18 sec.

**Scenario A: The MN (1.2.1) is connected to FA (1.3.0)**
As can be seen in Figure 4, the handoff delay for the proposed technique and the one level up are about 0.046 sec at simulation time 16 sec, while the handoff delay for the original MIP is about 0.052 sec at the same simulation time.
The packet loss is given in Figure 5, in which the number of packet losses during the handoff for the proposed technique is less than the one level up and the original MIP.

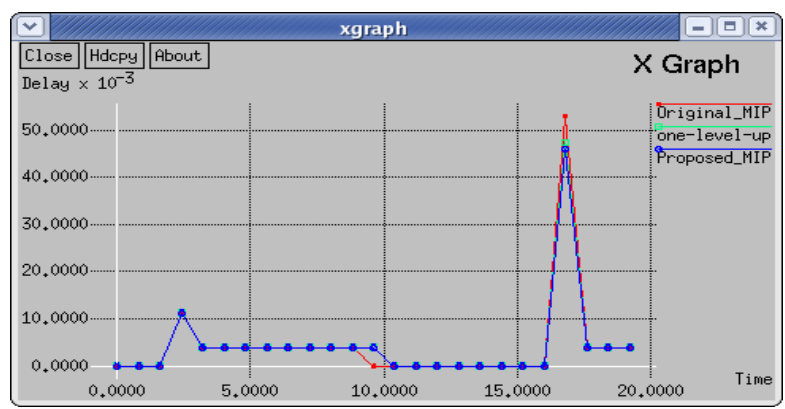

Figure 4. Handoff delay Comparisons of Scenario A

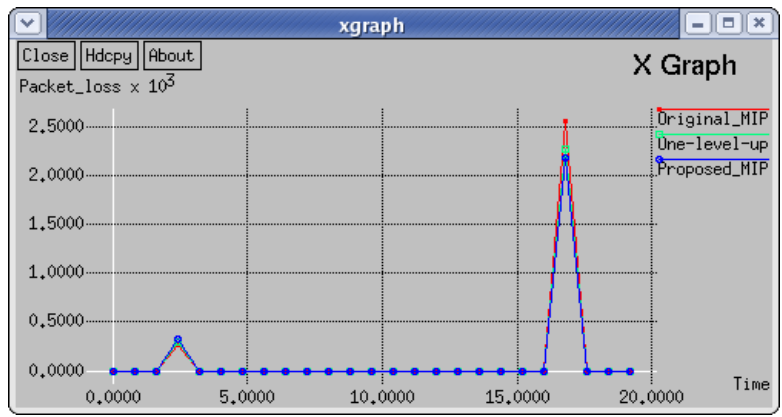

Figure 5. Packet loss Comparison of Scenario A

**Scenario B: The MN (1.2.1) is connected to FA1 (1.5.0)**
As illustrated in Figure 6, the handoff delay for the proposed technique, one level up, and the original MIP are about 0.062 sec, 0.065 sec and 0.070 sec respectively. These values occur at simulation time 16 sec. As in scenario A, the number of packet losses during the handoff for the proposed technique is less than the one level up and the original MIP techniques, as shown in Figure 7.

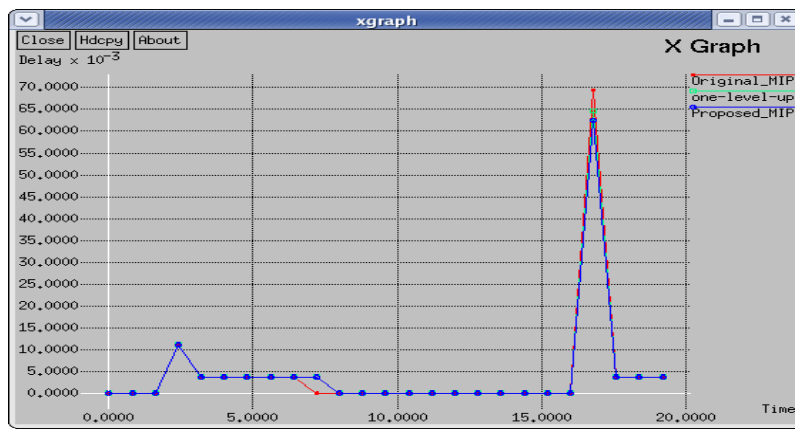

Figure 6. Handoff delay Comparisons of Scenario B

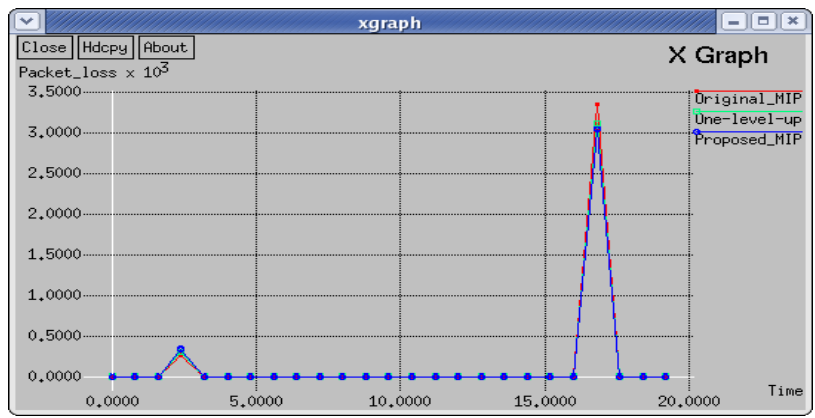

Figure 7. Packet loss Comparison of Scenario B

**Scenario C: The MN (1.2.1) is connected to FA2 (0.2.1)**

This scenario considers the case of existing the MN in a different domain. As can be seen in Figure 8, the performance of the proposed technique is better than both the one level up and the original MIP. The handoff delay for the proposed technique is about 0.061 sec and for the one level up is about 0.069 sec at simulation time 16 sec, while the handoff delay for the original MIP is about 0.072 sec at the same simulation time.

The packet loss is given in Figure 9, in which the number of packet losses during the handoff for the proposed technique is also less than the one level up and the original MIP.

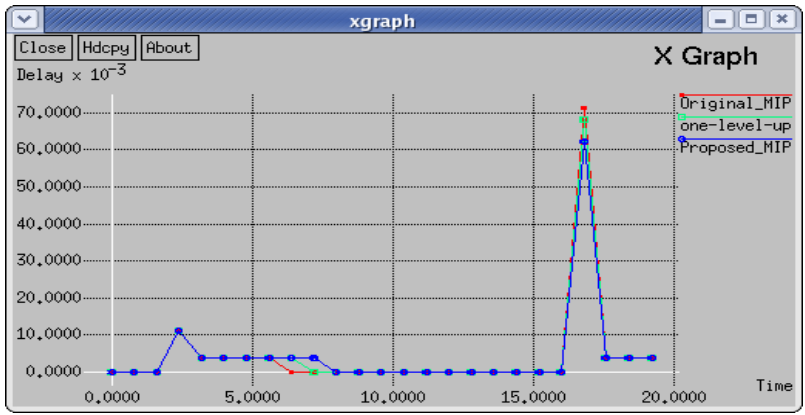

Figure 8. Handoff delay Comparisons of Scenario C

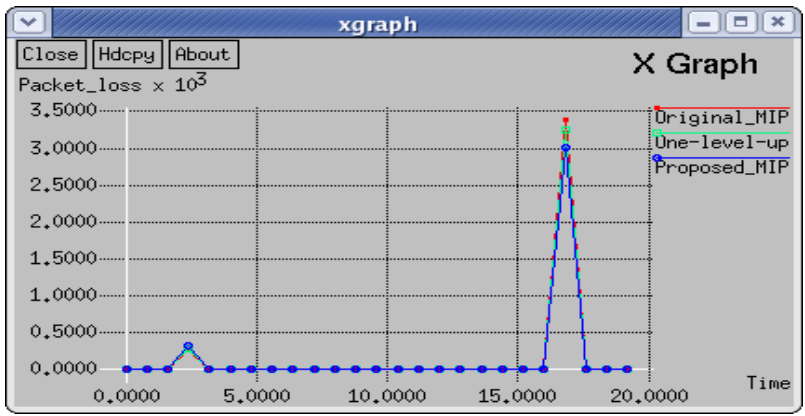

Figure 9. Packet loss Comparison of Scenario C

**Scenario D: The MN (1.2.1) is connected to FA3 (0.1.0)**

This scenario illustrates the case of existing the MN in a different cluster in the domain considered in scenario C. As can be seen in Figure 10, the handoff delay for the proposed technique is about 0.04 sec and for the one level up is about 0.053 sec at simulation time 16 sec, while the handoff delay for the original MIP is about 0.060 sec at the same simulation time. The packet loss is given in Figure 11, in which the number of packet losses during the handoff for the proposed technique is also less than the one level up and the original MIP.

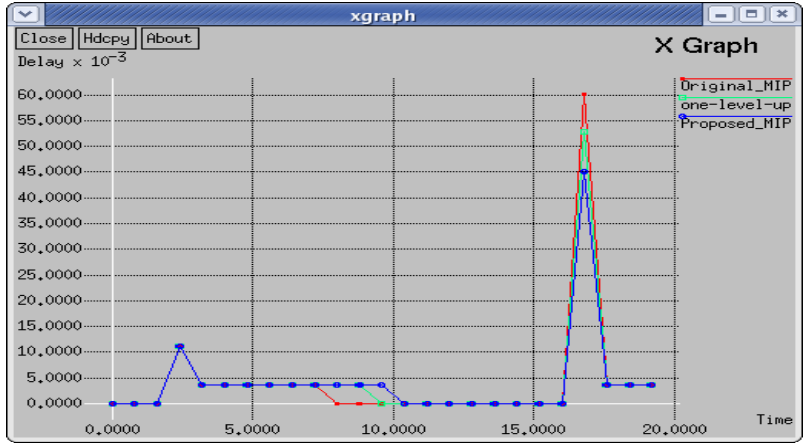

Figure 10. Handoff delay Comparisons of Scenario D

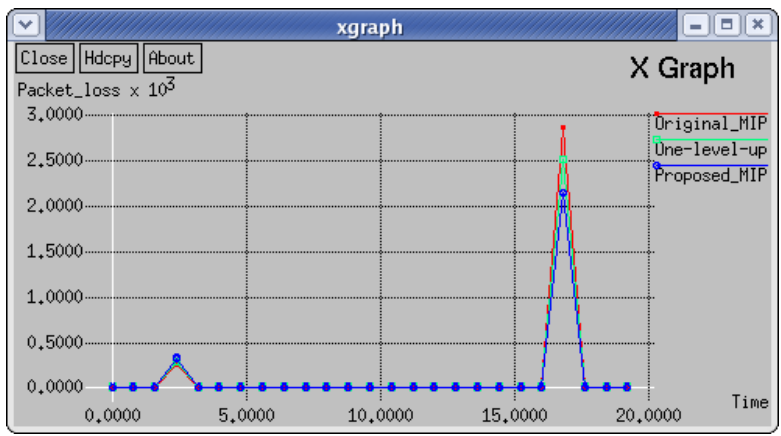

Figure 11. Packet loss Comparison of Scenario D

**Scenario E: The MN (1.2.1) is returned to it's HA (1.2.0)**

In this case, the handoff delay are almost the same for the proposed technique, the one level up technique, and the original MIP and it is about 0.11 sec at simulation time about 2 sec as illustrated in Figure 12. The number of packet losses during handoff for the proposed technique is much lower compared with both the one level up and the original MIP techniques as shown in Figure 13.

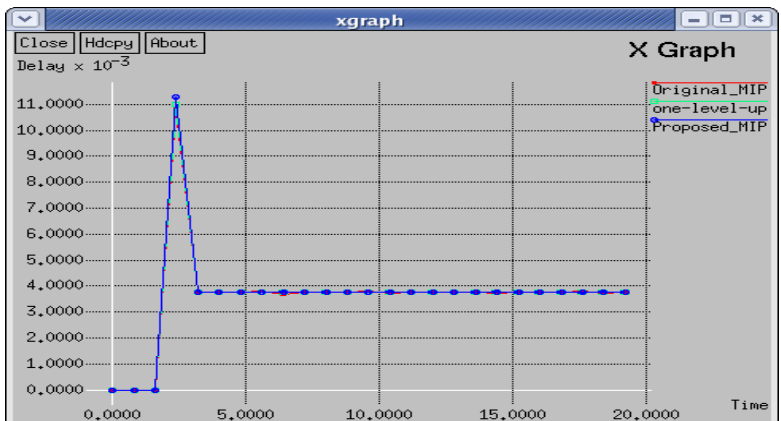

Figure 12. Handoff delay Comparisons of Scenario E

71

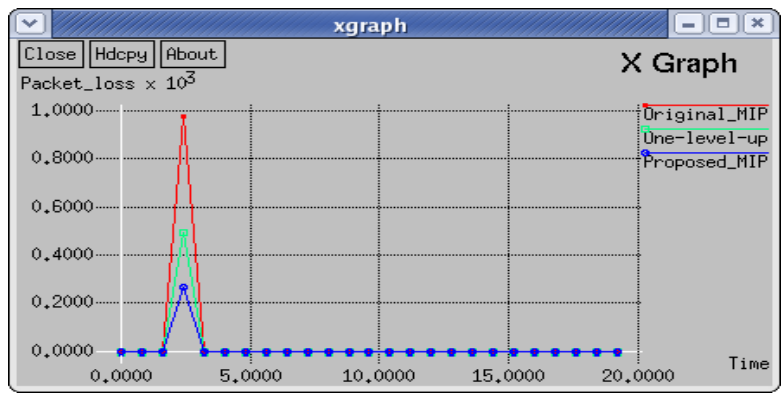

Figure 13. Packet loss Comparison of Scenario E

The other parameter considered in the simulation experiments is the registration time. The registration time is the time the MN takes to register and make conversation with the FA. Comparisons for registration time for the original MIP, the one level up and the proposed technique are presented in tables 4 and 5. The first column in tables 4 and 5 contain the address of FA serves the MN. As can be seen in these tables, it is clear that the registration times for the proposed technique are generally better than the other two techniques.

Table 4. Registration time values (in sec) for the original MIP and the proposed techniques

| Through FA | Original MIP | Proposed Technique | % Improvement |
| --- | --- | --- | --- |
| 1.5.0 | 0.808378 | 0.406199 | 49.75135395 |
| 0.2.1 | 1.206543 | 0.803992 | 33.36399946 |
| 0.1.0 | 0.808922 | 0.40611 | 49.79614845 |

Table 5. Registration time values (in sec) for the one level up and the proposed techniques

| Through FA | One level up | Proposed Technique | % Improvement |
| --- | --- | --- | --- |
| 1.5.0 | 0.606031 | 0.406199 | 32.97389077 |
| 0.2.1 | 1.00670 | 0.803992 | 20.13612754 |
| 0.1.0 | 0.60673 | 0.40611 | 33.06599949 |

## 5. CONCLUSION

In this paper, we proposed an efficient approach to deal with triangle routing problem in Mobile IP. Unlike the conventional MIP, tunneling process of the proposed approach occurs at two levels above the home agent in a hierarchical network instead of at home agent. Compared with other proposals, our proposal has primary advantages. The proposed mechanism can reduce packet delay, packet loss during handoff, and registration time. That is, unlike original MIP, a mobile node does not send any more registration update message to home agent after obtaining the new IP address. Instead, the location information for the mobile node is only updated at Surrogate Home Agent (SHA) (two levels above home agent). The simulation results, which are based on NS2 simulator, show that the proposed approach has a better performance compared to both the original MIP and the one level up technique. In future work, more experiments will be done to study the effect of increasing the levels above the home agent.